# Optics of an opal modeled with a stratified effective index and the effect of the interface

Isabelle Maurin[(1,2)], Elias Moufarej[(1,2)], Athanasios Laliotis[(1,2)],

Daniel Bloch[(2,1)]

[1] *Laboratoire de Physique des Lasers, Université Paris 13, Sorbonne Paris-Cité, F-93430 Villetaneuse, France*
[2] *CNRS, UMR 7538, 99 Avenue J-.B. Clément, F-93430 Villetaneuse, France*
*\*Corresponding author: isabelle.maurin@univ-paris13.fr*

**Reflection and transmission for an artificial opal are described through a model of stratified medium based upon a one-dimensional variation of an effective index. The model is notably applicable to a Langmuir-Blodgett type disordered opal. Light scattering is accounted for by a phenomenological absorption. The interface region between the opal and the substrate -or the vacuum- induces a periodicity break in the photonic crystal arrangement, which exhibits a prominent influence on the reflection, notably away from the Bragg reflection peak. Experimental results are compared to our model. The model is extendable to inverse opals, stacked cylinders, or irradiation by evanescent waves.**

## 1. INTRODUCTION

Structures with a refractive index periodically varying on a scale comparable to an optical wavelength are described as "photonic crystals" [1], and photonic energy bands can be predicted. The interest in photonic crystals is usually connected to their "bulk" optical properties, with the expectation of specific propagation rules *inside* the periodic arrangement of the crystal. Apart from the band structure, the photonic crystal periodicity (on a wavelength scale), and their possible self-organization, make them interesting in themselves. In nearly all cases, the optical information of interest requires a detection *outside* the photonic crystal. This makes the effect of the interface critical: (i) as a principle, this boundary region tends to break the periodicity typical of the photonic crystal; (ii) optical reflection at the interface usually probes the optical response on a dimension related to the optical wavelength λ. For these reasons, experimental tests of the quality of a photonic crystal are at a risk to explore mostly the quality of the first layers only, notably when the tests are conducted in reflection (see *e.g.* [2-3]).

Many techniques were developed to solve the difficulty of fabricating three-dimensional or two-dimensional photonic crystals [4]. Soft chemistry methods [5-10] are convenient in spite of a limited choice of geometry and materials. In particular, artificial opals can be obtained through the self-organization of a mono-disperse distribution of sub-micrometric spheres (in polystyrene, silica, or even $TiO_2$). In a process of sedimentation or convection involving evaporation of a solvent [5-8] and deposition of the spheres onto a substrate, the microscopic description of the evaporation of the solvent justifies that artificial opals, produced as a bulk material, tend to organize as a compact face-centered cubic (*f.c.c.*) crystalline arrangement [8] of dielectric spheres. The three-dimensional organization can be observed for good quality opals [11-15], through specific signatures associated to the simultaneous organization along the (111) and (200) crystalline planes. An alternate fabrication technique is the Langmuir-Blodgett (LB) method, which consists of a successive layer-by-layer deposition of spheres and allows a control of the number of layers [9, 10]. In such a technique, the photonic crystal, whose quality is affected by the sphere size dispersion, is usually highly polycrystalline, notably in the direction normal to the LB deposition plane: the crystalline arrangement usually exhibits a random hexagonal close-packed (*r.h.c.p.*) structure (i.e. a mix of *f.c.c.* and *h.c.p.* hexagonal close-packed lattices rather than a true *f.c.c.* arrangement). In all cases, the opal is not self-supported, but has grown from a (planar) substrate, with the (111) plane parallel to the substrate. The present work is devoted to a simple description of the optics of an artificial and rather disordered (LB-like) opal, and focuses on the influence of the interface region between the opal and its substrate.

Bandgap calculations solely based upon the opal periodicity (and indices, sphere diameter...) are not sufficient because the interface breaks the periodicity. Rather, various *ad hoc* models were developed in the literature. Numerical calculations based upon finite element methods (see [16]) can predict reflection, transmission, and also scattering including speckle and coherent multiple diffraction. They are nonetheless heavy, and of a limited versatility under a small change in the parameters of the opal or to deal with the randoms defects of a self-organized material. Alternately, simplified models based upon the "scalar wave approximation" (SWA) were also considered, relying on the periodicity of the photonic crystal (normally to the deposition plane); they are often restricted to normal incidence [17-20].

A fully homogeneous "effective index" model has also been popular in the photonic crystal literature [21] (and for opals since [6]), which restricts to a "zero-dimension" the crystal periodicity. In this approach, the concept of "effective index", initially developed in optics to take into account inclusion of "small-size" crystallites (i.e. small relatively to the relevant optical wavelength), is extended to glass spheres whose size is of the same order of magnitude as the wavelength. This crude model has even been used to describe antireflection properties of a single layer opal [22, 23]. It is actually often combined with the geometrical periodicity of the opal crystalline arrangement [6, 11-13, 24-25] in order to predict a peak of reflectivity in analogy with a Bragg diffraction peak, so that the "effective index" allows to determine the optical periodicity inside the opal. By adjusting the model to experiments performed under various incidences, a refined estimate of the sphere diameter and/or of the sphere index (i.e. an indication of porosity) is sometimes obtained [13, 26]. However, such an approach cannot predict quantitatively transmission or reflection (nor scattering).

In the present work, we consider a stratified one-dimensional version of an effective index model. This stratified description of the opal layers allows a much better description of the interfacial regions between the opal and the substrate (or the vacuum), notably for these extreme regions where the glass spheres are not in a compact arrangement. Such a model, although mentioned in some occurrences in the literature [27, 28], has not been fully developed until now [29]. Among several advantages, including rather simple and versatile one-dimensional calculations, it appears well-suited to a thin LB opal with its already mentioned *r.h.c.p.* structure. The purpose of such a one-dimensional index model is intrinsically restricted to a fair description of reflection and transmission behaviors, while light scattering and diffraction is out of reach. Rather, an *ad hoc* loss [28] has to be added to the model to avoid the sum of the reflection and transmission coefficients to be unity for a transparent material like glass. The major original result of our model is that it demonstrates the strong influence of the opal "interface", which is defined, depending on the experimental conditions, as the contact region including the substrate on which the opal is deposited, or as the non planar interface between vacuum and the opal. This single interface is at the origin of most of the reflection properties, notably for all incidences falling far away from the specific condition of Bragg reflection. We show also that the model, which intrinsically includes the complex interferences between all thin slices of opal, applies satisfactorily to understand the build-up of a quasi "Bragg diffraction" and the residual Fabry-Perot-like oscillations [5, 11-12, 15, 30], or the effect of an imperfect opal periodicity [11, 30]. The extension of our results to the specific resonant properties of a material infiltrated in the void regions of an opal [31] has also been considered, and is the topic of a specific paper [32]. At last, results derived from some of our previous experiments [33], which were mostly devoted to the analysis of the infiltration of a resonant gas into an opal, and from complimentary dedicated experiments on some smaller diameter opals, are shown to be compatible with our predictions once the extinction parameter is adjusted.

The paper is organized in the following way: in section 2, we describe the opal as a stratified medium, and introduce the treatment through a transfer matrix approach. Although the transfer matrix is a classical method, we provide in Appendix A some details to validate the introduction of an absorption coefficient replacing scattering. In section 3, we discriminate the contribution of the different regions of the opal, through simplified distributions of the stratified index (e.g. "fused opal" model), and notably discuss the specific influence on reflection of the interface between substrate (or vacuum) and the opal. The coherent construction of a Bragg reflection associated to the (bulk) periodicity of the opal, and the effect of a single layer defect introduced on purpose or as a fabrication defect, are also considered. The next section (section 4) first summarizes our previous experimental findings concerning the reflection at an opal interface at specific $\lambda/D$ values ($\lambda$ the optical wavelength, $D$ the diameter of the opal spheres) and reports on dedicated experiments performed with a white laser on small size opals (with numerous defects), in a regime making the Bragg reflection observable. The conclusive section (section 5), summarizes the results and briefly considers some possible extensions.

## 2. OPTICAL MODEL OF THE OPAL AND FORMALISM

### A. Stratified effective index

The "effective index" model has been generally applied to average, through a single parameter, the optical properties of a medium that includes voids or impurities, whose size is much smaller than the optical wavelength ($\lambda$) (see e.g. [34] and refs. therein). The effective index $n_{eff}$ is deduced from an "averaged" permittivity. For a dielectric medium with small size (vacuum) voids, it is hence defined as:

$$n_{eff} = [f.\varepsilon + (1 - f)]^{1/2} \quad (1)$$

with $\varepsilon$ the dielectric constant of the filling material ($\varepsilon = n^2$ with n the dielectric index of the filling material), and $f$ the (volume) filling factor of the dielectric medium.

To extend an "effective index" description to a "stratified effective index" model for the opal, one has to "slice" the opal in successive layers parallel to the (111) plane. In the case of a LB opal, many defects may occur for each deposited layer, but the overall structure of successive deposited layers is respected as long as a cumulative disorder has not totally washed out the crystalline structure. This is the reason to apply a model for which the "effective index" $n_{eff}(z)$ is stratified along the direction $z$ perpendicular to the substrate, with the density measured only over a plane of an opal layer, parallel to the substrate. Although slicing is essentially a discrete operation, there is no inconvenience in going to a continuous description, with infinitely thin slices. This requires to consider a spatially-dependent (along $z$) filling factor $f(z)$ -in the plane- of the dielectric material (circles for opal). The stratified effective index distribution is hence given by:

$$n_{eff}(z) = [f(z).\varepsilon + (1 - f(z))]^{1/2} \quad (2)$$

as long as the medium surrounding the sphere is assumed to be vacuum, and $f(z)$ is discussed below.

### B. Sphere packing in an ideal opal and periodicity index

We consider here the case of an ideal opal, in order to use the resulting $f(z)$ in eq. (2). The ideal opal is a close-packed arrangement of identical spheres, which can be described as successive layers of bi-dimensional sections of close packed spheres (a single layer of spheres). The successive layers themselves are arranged in a compact manner, so that the distance between two successive layers is $(2/3)^{1/2}.D$, with $D$ the sphere diameter.

For the first layer above the substrate (which we locate in the $z \leq 0$ region), the filling factor, which corresponds to a closed-packed distribution of circles in hexagonal cells in the equatorial plane ($z = D/2$), is given by:

$$f_1(z) = 2\pi z.(D-z)H(z) / (3^{1/2} D^2) \quad (3)$$

with $H(z) = 1$ for $0 \leq z \leq D$ and $H(z) = 0$ elsewhere.

For an opal made of N layers arranged in a compact manner, the filling factor $f(z)$ has to take into account the interpenetration between layers, and is the sum of the filling factor of the individual layers:

$$f(z) = \sum_{i=1}^{N} f_1 \left[ z - (i-1).D.(2/3)^{1/2} \right] \quad (4)$$

In Fig. 1, we show the filling factor for an opal of glass spheres (fig. 1a), and the corresponding effective index -in a discretized slice version- (see fig 1b). The two structures are very similar, with the $(2/3)^{1/2}.D$ periodicity between the first equatorial plane (at $D/2$) and the last one. Figure 1 clearly shows that the "opal and substrate" system combines an internal region with periodicity (here $(2/3)^{1/2}.D$ because we have assumed a compact opal), and nearly empty regions (of a thickness $D/2$) close to the substrate. The consequences of this periodicity break, occurring in the interfacial regions (for the first and last half-layers as well), are discussed in section 3.

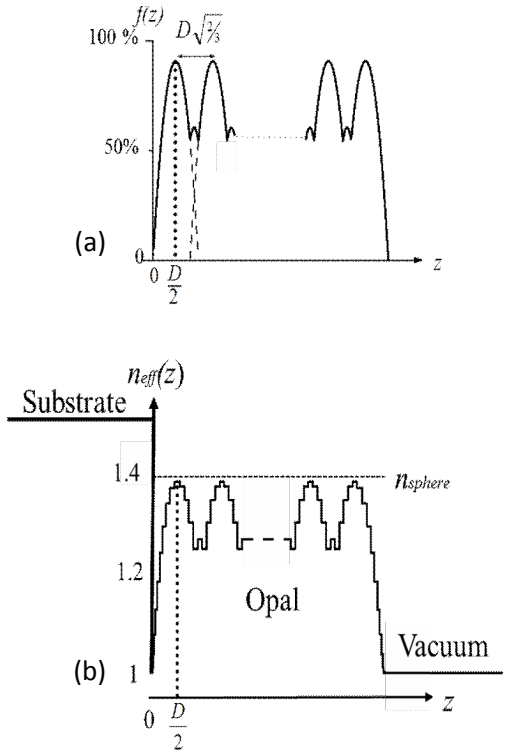

Fig 1: (a) full line: the filling factor $f(z)$ in the $z$-height plane for an opal made of N layers. The dashed lines below $f(z)$ represent the individual filling factors for the first and second layers; (b) The one-dimensional effective index $n_{eff}(z)$ of the opal, for $n_{sphere} = 1.4$, shown here in discrete version (stratified medium made of finite thickness layers)

### C. Propagation in a stratified medium

#### 1. Matrix formalism and reflection/transmission coefficients

The matrix formalism [35] for a stratified medium made of successive parallel transparent layers allows calculating the reflection and transmission coefficients. The elementary matrix allows going from one layer to a neighboring one, with the limiting boundary conditions provided by the continuity of the tangential components of the total (electric and magnetic) field at an interface, and Snell's law for transmission and reflection have to be satisfied at the successive interfaces. In this formalism, the two principal polarizations, TE and TM, do not mix, and have to be dealt with separately. For a real opal, this implies that we assume that principal polarizations do not mix through propagation, although such a mixing can be predicted when taking into account all the symmetries of a f.c.c. opal [36].

The matrix formalism can also apply in the presence of absorption losses [see Appendix A, notably eq. (A18)]. As mentioned in section 1, we introduce such a phenomenological loss in lieu of scattering. Let us note that the absorption integrated over a single layer must remain small for the stratified model to be a useful description of the opal structure: indeed, if the scattering on a single layer of spheres would become too strong, it would be difficult to identify an effect of the crystalline organization of the opal along $z$. Note also that according to Snell's law, the incidence angle to be considered for a given layer exhibits a complex value, which no longer represents a direction of propagation [37, 38]. However, as long as the input and output media (vacuum or glass) are assumed to be transparent, there is no difficulty in defining the incidence, reflection, and emergence (real) angles. Moreover, the change in the propagation direction is negligible when the loss term in the complex refractive index remains small. The corresponding details are provided in the appendix A.

#### 2. Input and output media for an opal deposited on a substrate

We consider an opal which is deposited on a transparent substrate, like a glass (parallel) window. Two situations of interest can be discriminated: light entering from the substrate side (as in our experimental work [33] with a resonant gas), or from vacuum (or air). In all cases, the media outside the opal are transparent, and they are considered to be infinite in the matrix formalism. Actually, in an experiment, the transparent substrate has a finite size and light first propagates in the vacuum before entering into the window. For a thick enough window, a minor wedge angle, which does not affect the direction of light propagation inside the opal, is sufficient to make negligible all internal interferences inside the window, and the specific transfer matrix associated to the (thick) window can be ignored. The effect of this slab is equivalent to the one of an "incoherent" layer [37]. For convenience, to help comparing the cases when light enters from the substrate or from the vacuum region, we use in all the following the "external" incidence angle, labelled $\theta$. This "external angle" is defined as the angle in vacuum: this assumes that when light goes from the substrate (index $n_0$) to the opal (with an incidence $\theta_0$ in the substrate), the substrate itself is illuminated from the vacuum, under an "external" incidence angle $\theta = \sin^{-1}(n_0 \sin\theta_0)$.

Note that our assumption that the substrate on which the opal is deposited is a nearly parallel slab (index $n_0$) implies that $n_0 \sin\theta_0 \leq 1$. A larger range of effective incidence angles, not restricted to $\sin\theta_0 \leq 1/n_0$, can be considered by our formalism, and would be very interesting to probe. Experimentally, such a situation could be addressed in a total reflection geometry for an opal deposited on a prism-shaped substrate, allowing the probing of the attenuated reflection as induced by the transmission to the opal.

#### 3. Loss coefficient equivalent to the scattering effect

It is a known and remarkable point that for a slab (or layer in a stratified medium) of a given thickness, the dephasing associated to propagation decreases when the incidence angle increases. Conversely, as recalled in the appendix, the field attenuation increases with the incident angle proportionally to the length traveled inside the medium. This says that the loss coefficient to be introduced may have to depend on the incidence angle. This phenomenological absorption should also depend on the wavelength, through the sphere size/ wavelength ratio $D/\lambda$ (or $Dn_{sphere}/\lambda$) and possibly on the polarization (TE or TM).

In most cases, we neglect the incidence angle dependence mentioned above, in rough agreement with experimental behaviors that we report on in section 4. We have mostly performed our numerical simulations with a loss coefficient independent of the wavelength, or sometimes following a kind of Rayleigh law in $\lambda^{-4}$.

The formal calculation of a stratified medium easily allows for a spatial distribution of the loss $\alpha(z)$ distribution. In various occurrences, we have compared a spatially constant loss model [$\alpha(z) = \alpha$], with a model where losses occur only inside the sphere [$\alpha(z) = \alpha f(z)$], or in the voids [$\alpha(z) = \alpha (1 - f(z))$], or where the losses depend on the contact surface [*i.e.* on the sphere perimeter in the considered layer $\alpha(z) = \alpha f(z)^{1/2}$]. As long as the average (small) loss per layer remains unchanged, there are no significant changes to be predicted when varying the model taking into account the scattering losses. It is hence just simpler to consider a constant loss model.

## 3. MAJOR RESULTS OF THE MODEL

### A. General behavior of the model and typical numerical values

To predict the optical behavior of an opal deposited on a substrate, we apply the stratified model (section 2 and appendix) with numerical values chosen according to the experimental conditions, polarization (TE, TM, or linear combination), the side of incidence (opal or substrate side), incidence ("external") angle $\theta$, and irradiation wavelength $\lambda$. For the opal, the relevant figures are the number of layers (N) - typically up to 10 or 20, as scattering tends to hinder the role of the deepest layers -, the sphere index $n_{sphere}$ (in all the following, we take numerically $n_{sphere} = 1.4$) from which $n_{eff}(z)$ is deduced, the substrate index $n_{substrate}$, the sphere diameter $D$ -or only the reduced parameter $D/\lambda$ -, and the absorption $\alpha$, possibly dependent on $\lambda$ or on $\theta$.

Numerically, the calculations for a given opal are performed for a finite number of slices and converge when this number is increased. Practically, it is efficient to divide the opal in a given number ($N_{step}$) of layers - of equal thickness- per period in the periodic region, and in $N_b$ steps to describe the sharper variations of $n(z)$ on the first and last half-layers. For a faster convergence, these latter "slices" in half-layers regions have unequal thicknesses in order to ensure a regular growth for $n(z)$. In these conditions, when $D$ and $\lambda$ are comparable, we avoid any convergence problems by taking $N_b \sim 50$, and $N_{step} \sim 20\text{-}40$, meaning that a total of less than $10^3$ steps is needed for a rather thick opal made of 20 layers.

To illustrate several of the typical behaviors that we will detail in further subsections, we present in figure 2 a reflection spectrum calculated using the model described in section 2, covering a considerable range of wavelength (or $\lambda/D$ value). We assume an opal made of a large number of layers (N = 20) of glass spheres (with $n_{sphere} = 1.4$ as usual), deposited on a similar glass substrate ($n_{substrate} = 1.4$). Either we ignore the scattering losses (fig. 2a), either we replace scattering by an absorption independent of $\lambda$, which remains small enough for a single layer, but is non negligible for the whole opal (fig. 2b) [39]. Differences are only quantitative between these two reflection spectra. Major similarities are found, showing that the detailed description of the scattering loss is not essential. Note that the equivalent transmission spectra would be strongly affected by the choice of the loss model and this is why all discussions in the present section 3 will be restricted to reflection.

Figure 2 permits to distinguish several important features. First, one recognizes a sharp reflection "Bragg peak" for $\lambda/D \sim 2.06$. As it is well-known, this strong peak of reflection may mean that light cannot enter into the opal, as this situation corresponds to a bandgap region of a photonic crystal, even if the opal is not a fully periodical system. When absorption is taken into account (fig. 2b), light is only partly reflected, and partly lost inside the opal: the system of destructive interferences that prohibits propagation has become imperfect. The detailed features of this reflection peak (width, asymmetry, lineshape,...), are essentially determined by the numerical values and shape modelling chosen for the opal. This is to be compared with numerous studies in the literature where the lineshape of the Bragg peak is just described by a quantity such as a FWHM (full width half-maximum) (see *e.g.* [11]), ignoring the asymmetry of a real lineshape.

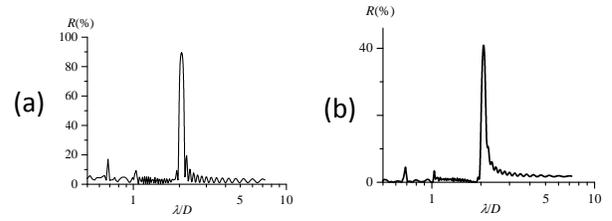

Fig 2: Reflection spectrum calculated for N = 20 layers. Irradiation on the vacuum side, TE polarization, $\theta = 20°$, $n_{substrate} = 1.4 = n_{sphere}$. The absorption coefficient is: (a) $\alpha D = 0$; (b) $\alpha D = 0.1$

Two other peaks of a smaller amplitude can also be identified ($\lambda/D \sim 1.04$ and $\lambda/D \sim 0.7$), which are known to be associated to high-order Bragg diffraction. Apart from these peaks, reflection remains rather weak and exhibits recognizable oscillations. All these features, from small oscillations to the Bragg peaks asymmetry, originate in interference effects.

In the next subsections, we address the major features of Fig.2, replacing when possible the stratified opal by even simpler models, to focus on a specific effect. In particular, we discriminate the essential effect of the periodicity break in the vicinity of the interface region, from various effects of periodicity (Bragg peak, and Fabry-Perot like effects) already well-identified in the literature. We also include a local defect inside the opal [40-42]. For simplicity, we choose the loss coefficient $\alpha$ to be wavelength-independent in this section 3.

### B. Effect of the interface between the opal first half-layer and substrate or vacuum on reflectivity

It has been recognized that reflectivity of a photonic crystal is not always a proof of its good organization, and even that the reflectance of a "photonic glass" [2], which is an intrinsically disordered medium made of identical spheres, may resemble the reflectance of the equivalent photonic crystal.

Here, to outline the symmetry break imposed by the interface region, rather than the effect of opal periodicity, we consider a specific stratified system of a "fused opal", whose index varies according to the local effective index (eq. 2) for the first half-layer of spheres ($0 \leq z \leq D/2$), and where the periodic region of the opal is replaced by a "fused" region of constant index $n_{max}$ (on an infinite length, or at least, on a length largely exceeding the extinction length, see fig. 3a). For continuity reasons, we assume $n_{max} = n(z = D/2)$, so that we choose numerically $n_{max} = 1.36$ (for $n_{sphere} = 1.4$). We take a moderate absorption coefficient $\alpha D = 0.1$ (see Fig 3) to allow a finite-size of the "fused opal".

The reflection coefficient is expected to depend on the ratio $\lambda/D$ which governs the ratio between the wavelength and the extension ($\sim D/2$) of the "gap" region. Two limits can be predicted : for $\lambda/D >> 1$, the "gap" region becomes so thin that it can be ignored and the reflectivity can be simply estimated -through Fresnel formulae- from the interface between the substrate (or vacuum) and the constant index of the "opal" with constant index; conversely, for short wavelengths $\lambda/D << 1$, light mostly feels the contrast between the substrate and vacuum at the contact plane with the spheres (or a null contrast for an irradiation from the vacuum region), and further accommodates with the slowly varying index. Here, we provide an insight of the predicted evolution between these two extreme behaviors, covering a large range of $\lambda/D$ ratios. Figures 3 and 4 summarize the results for the two principal polarizations and various incidence angles.

In fig.3 (b,c), reflectivity on the vacuum side (*i.e.* $n_{substrate} = 1$) in the short wavelength limit is extremely small as expected, and increases in a nearly monotone manner with increasing $\lambda/D$ ratio. For $\lambda/D \to \infty$, it typically approaches the Fresnel reflection coefficients (which are incidence and polarization dependent) at a vacuum/$n_{max}$ interface.

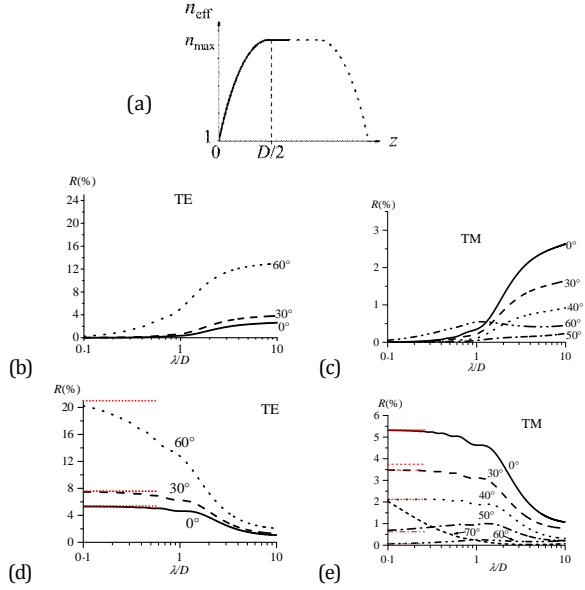

Fig 3: Reflectivity for the "fused opal" model: (a) effective index of the "fused opal"; (b,c,d,e): normalized wavelength ($\lambda/D$) dependence of the reflectivity for the indicated external incidence angle $\theta$. The irradiation is: (b,c) from the vacuum side; (d,e) from the substrate side. Polarization is TE for (b,d) and TM for (c,e). In(d) and (e), the dotted or dashed lines (red on line) for the smaller values of $\lambda/D$ indicate the Fresnel reflection coefficient for a planar substrate/vacuum interface. Calculations for $\alpha D = 0.1$ and for (d) and (e), $n_{substrate}$ = 1.6.

For reflection on the substrate side, we choose a high value $n_{substrate}$ = 1.6 in Fig. 3 (d,e) to show how one goes from a Fresnel reflection limit from "$n_{substrate}$/vacuum" to a "$n_{substrate}/n_{max}$" interface when going from the short wavelength limit (i.e. long "gap" region) to the long wavelength limit (i.e. short "gap" region). The same kind of nearly monotone evolution appears in Fig. 4 showing the influence of the substrate index (calculated for the normal incidence). In TM polarization (see Fig. 3 c,e), the reflection coefficient remains very small around 50-60° incidence, a behavior reminiscent of a Brewster angle at a single flat interface. In spite of these simple trends, one also observes, when looking more in detail on figs 3 and 4, tiny wavy behaviors. They are a signature of the multiple interferences intrinsic to a stratified index medium, which is here restricted to a "smooth" and monotone half-layer interface.

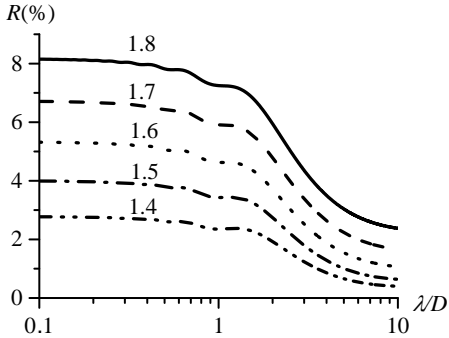

Fig 4: Normalized wavelength ($\lambda/D$) dependence of the reflectivity for the "fused opal" model, for normal incidence irradiation from the substrate side. Values of $n_{substrate}$ as indicated. Calculations are for $\alpha D = 0.1$.

From figs 3 and 4, one sees that the behavior for an irradiation on the substrate side (figs. 3 and 4) remains close to the asymptotic regime of small $\lambda/D$ values, where the dominant effect is the "gap" between the substrate and the thin empty region, as long as $\lambda \leq D$. It is rather for very long wavelengths, when the effect of dephasing (~ 1 rad, i.e. propagation over $\lambda/2\pi$) cannot occur on a distance comparable to the interface thickness ($D/2$), that the gap region can be neglected. It is worth noting that around the "Bragg reflection peak" evidenced in Fig. 2, which is found for $\lambda/D \sim 2$, the effect of the interface region is not yet negligible, and must be fully taken into account for a quantitative description of the amplitude or shape of the Bragg peak.

### C. Bragg reflection and Fabry-Perot oscillations

In Fig. 2, the sharp peak for $\lambda/D \approx 2.06$ in the reflectivity spectrum is a well-known feature of an opal [6, 9, 11-13]. It is associated to a "forbidden band" in a photonic crystal approach and corresponds to a nearly prohibited transmission. Describing this strong reflectivity as a "Bragg reflection" is actually a way to underline an analogy with the X-ray diffraction on the periodically located point-like nuclei in the structure of an atomic or molecular crystal. Note that the extremely small size of nuclei, and their precise arrangement relatively to the distance between successive planes, lead to a definition of the Bragg angle which is incommensurately sharper than the one encountered with opals. For an opal, the periodic three-dimensional grating actually relies on an ensemble of contacted spheres, but the principle of a geometric condition governing the direction of a Bragg diffraction still applies. The standard Bragg condition $k\lambda = 2a\cos(\theta)$, with k an integer, $a$ the distance between successive planes, and $\theta$ the incidence angle, should however include the effect of propagation in the opal as an heterogeneous medium, so that a modified equation [6] is often applied to find the wavelength $\lambda_{max}$ for an opal [12-13, 24]:

$$k\,\lambda_{max}/D = 2.(2/3)^{1/2}\,(n^2_{eff} - \sin^2\theta)^{1/2} \qquad (5)$$

Eq. (5) takes into account the $(2/3)^{1/2}.D$ geometrical periodicity of the opal, and the Bragg condition is satisfied in a fictitious medium whose index is the averaged effective index $n_{eff}$ (usually estimated with the compact filling factor 74 %), and for which the effective incidence deduces by the Snell's law from the external incidence angle $\theta$ (i.e. in vacuum). The peak at $\lambda/D \approx 2.06$ in fig. 2 is hence compatible with a first-order (k=1) Bragg diffraction peak.

The sensitivity of the Bragg peak region to the details of the opal shape is illustrated in fig. 5. Figure 5 is the result of a calculation for a simplified opal, in which the smooth "gap" regions (first and last half-layers) are replaced by a single arbitrary step ($N_b=1$). The behavior is analogous to the one shown in fig.2 around the Bragg peak. However, the residual lineshape asymmetry around the Bragg peak differs from the one appearing in fig. 2. The lineshape is again modified, even with minor changes in the shape of the periodic structure, or when keeping a detailed description ($N_b$ = 50) of the interface region as it is the case in fig. 6 (see e.g. for $N_{layer}$ = 10 or for $N_{layer}$ = 30).

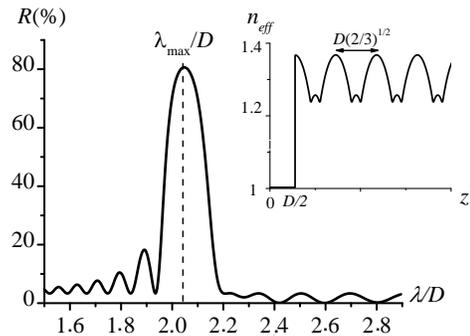

Fig 5: Reflectivity spectrum for a periodic distribution of the effective index $n_{eff}(z)$ as shown in the inset. Calculations are for N=18, $\theta$= 20°, $n_{substrate}$=1.4, $\alpha D$=0.193, and TM polarization.

Also, it is when increasing the number of layers that the contribution of the periodic region can be recognized. One typically needs at least 3 layers (see fig. 6) to recognize a Bragg diffraction peak in reflection. The peak gets sharper (maximized amplitude and minimized width) when increasing the number of layers, while the position of the peak moves to an asymptotic value, obtained for ~10-20 layers with our realistic choice of parameters. The number of layers needed to reach the asymptotic behavior of a thick opal depends on the scattering/extinction parameter.

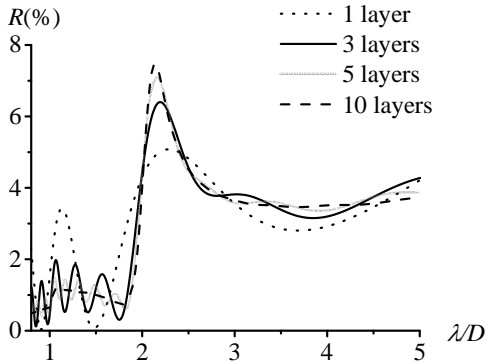

Fig 6: Reflectivity spectra (vacuum side), as a function of the number of layers N (as indicated). TE polarization, $n_{substrate}$ = 1.4, $\alpha D$ = 0.0552, $\theta$ = 20°. The curve for N = 30 (not shown) is just superimposed to the one for N = 10.

Our model also provides lineshape and amplitude predictions for the already mentioned higher-order Bragg reflectivity peaks, as can be revealed by some differences between fig. 2a and 2b. However, a practical limitation for the validity of these predictions is due to the absorption coefficient replacing the effect of scattering. This effect can indeed undergo considerable variations over an octave of wavelength.

Apart from the first and high-order Bragg reflection peaks, tiny quasi-periodic oscillations are seen in figs. 2, 5 and 6. One can note in fig. 6 that their periodicity and amplitude decrease when increasing the number of layers. These oscillations, which disappear in a too simple model when the exit layer is not reached, are associated to a Fabry-Perot like effect between the two extreme regions of the opal. As can be verified on fig. 6, they tend to cancel for a thick opal because of the extinction by scattering. These oscillations are often mentioned in the literature [11-12], and are sometimes used to evaluate the opal thickness or to assess the parallelism of the deposited layers.

### D. Effects of a defective layer in the opal

The quality of a real opal arrangement strongly depends on the uniformity of sphere size, which can be good for polystyrene spheres, while dispersion is higher for glass spheres. Also, the opal quality tends to degrade layer after layer due to dispersion in sphere size, clustering of spheres, sphere porosity, arrangement defaults, etc... This is a major limitation for all theories relying on the periodicity of the opal.

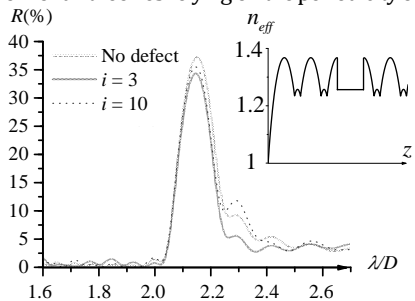

Fig 7: Reflectivity spectra (vacuum side) under normal incidence for various positions of the $i^{th}$ layer carrying a defect (see inset); $n_{substrate}$ = 1.4, $\alpha D$ = 0.09.

To illustrate the flexibility of our model, and to approach a realistic situation [20, 41] on the road to a disordered opal, we introduce an on-purpose defect in the $i^{th}$ layer of the opal, assuming for the index a sudden change to a constant value which locally breaks the periodicity. Figure 7 illustrates how the shape of the reflectivity spectrum changes with the position of the defect. Whatever the position of the defect layer, the peak reflectivity decreases, while the position of the Bragg peak is marginally modified. The change is stronger when the defect is located among the first layers. This had to be expected as the first layers contribution is critical for the build-up of the Bragg peak, while absorption (or optical scattering) makes the contribution of the remote layers less important, as already seen in figure 7.

## 4. EXPERIMENTAL OBSERVATIONS

### A. Initial observations

Our initial interest for the optics of opals started with spectroscopic experiments dedicated to the analysis of the resonant behavior of a gas infiltrated in an opal. These experiments, initially conducted at $\lambda$ = 852 nm and $\lambda$ = 894 nm on opals made of 10 or 20 layers of glass spheres with $D$ = 1.0 μm deposited on a standard glass ($n_{substrate} \approx$ 1.5), were enriched by experiments with $\lambda$ = 455 nm, and with opals $D \sim$ 400 nm [33]. In addition to the scattering of the incident light - whose residual coherence occasionally leads to specific diffraction figures, such as an hexagonal diffraction for a single opal layer [43], or to more complex structures for a perfect multilayer arrangement [11, 37] -, we had noted that a fraction of the beam undergoes a specular reflection at the substrate/opal interface, while another part of the beam is transmitted. The presence of such reflected beams, although well documented even far away from the Bragg reflected peak [33, 44], is partly unexpected because the opal/substrate and opal/vacuum interfaces are strongly non planar (at a microscopic/wavelength scale). Also, the incident beam polarization (assuming a principal polarization for the irradiation) is mostly conserved [45] for these reflected or transmitted beams. A major result is that reflectivity at the substrate /opal interface (for 10 or 20 layers as well) is extremely close to the one measured on the bare substrate (~ 4% under normal incidence). This is true independently of the incidence angle or polarization, as long as the reflection remains rather small (practically, this regime breaks only for TE polarization above 30-40 °). For large incidence angles and TM polarization, reflection on the opal becomes very small, a situation analogous to the Brewster incidence on a flat surface. All these behaviors appear fully compatible with the results of our one-dimensionally stratified model, when we show (section 3.B) that reflectivity at the interface remains close to the value predicted for a simple planar interface as long as $\lambda$ does not largely exceed the "gap" size.

### B. Dedicated experiments

In this section, we report on dedicated experiments performed on an opal where we used a white light irradiation, in order to test some of our predictions related to the Bragg-type reflectivity peak. Also, we measure both transmission and reflection, in order to explore how a simple conversion of the scattering effect into a "loss" term can be acceptable.

#### 1. Experimental set-up

Our experiments with white light were restricted to a single sample made of 20 layers of silica spheres smaller than in previous section ($D$ = 280 nm ± 5%). The Langmuir-Blodgett deposition technique was used to fabricate this opal, which has many structural defects owing to the dispersion in sphere size, resulting in a polycrystalline opal. The substrate of the opal is a microscope slide (glass), whose index is close to the one of the $SiO_2$ microspheres; the parallelism of the microscope slide is poor enough to make internal interferences negligible.

The sample was illuminated with a collimated white light from a fibered type supercontinuum source (LEUKOS SM 20, for 400-1700 nm, pulses < 1ns, repetition rate: 20Hz, average power > 40 mW). The beam diameter was around 2 mm, corresponding to an averaging on many spheres (at this scale, each layer of the opal is already polycrystalline). The beams of interest (the reflected one, and the transmitted one) are collected by a second optical fiber connected to a fibered spectrometer (Ocean Optics USB 2000+, detection for $\lambda$ = 200-1100 nm, resolution $\Delta\lambda \sim 0.4$ nm). The orientation of this detection fiber must be carefully optimized to ensure that the collection efficiency is the same all over the spectrum. This is an important source of uncertainty for a multimode fiber and for an analysis over a rather broad spectra range, and this sets limits to our experimental sensitivity. Despite these limitations, our goal is to compare effective measurements with predictions, and not just lineshapes in arbitrary units. Reflection and transmission spectra of the opal are obtained once normalized against the spectral content of the white source (temporal fluctuations are notable, requiring frequent analyses of the spectrum content of the white source). For sensitivity reasons, reliable spectra are obtained mostly in the 400-800 nm region, covering for our opal a 1.5 - 3 range of $\lambda/D$ values, complementary to our initial experiments.

The reflection and transmission spectra were recorded for various incident angles $\theta$, both for an irradiation on the vacuum side and on the substrate side. The incident polarization was controlled by an external polarizer. In our experiments, the polarization of the beam after reflection or transmission in the opal appears unchanged despite the three-dimensional scattering [45].

## 2. Transmission measurements and the evaluation of scattering losses

In the previous section (section 3), we used our model only for the opal reflectivity, rather than for transmission, because some important features of reflectivity are rather insensitive to the scattering losses. Transmission, intrinsically very sensitive to the absorption term, is also provided by the matrix formalism (section 2).

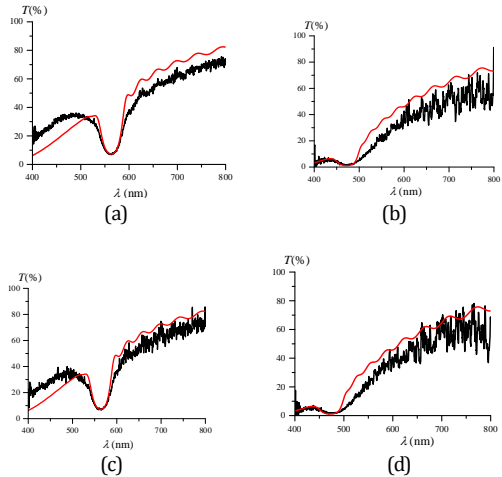

Fig. 8: Transmission spectra (TE polarization) on an opal made with 20 layers of $D$ = 276 nm glass spheres: (a,b) vacuum side; (c,d) substrate side; for (a,c) $\theta$ = 20°; for (b,d) $\theta$ = 50°. Experimental curves (in black) are compared to the calculated values (red on line), for which one has taken $\alpha = \beta/\lambda^4$, with $\beta$ = 1.5. $10^{-20}$ m$^3$.

Experiments (see fig. 8) clearly show a wavelength dependence for transmission, which drops down in the short wavelength region although glass remains transparent. One also notes a strong dip in the transmission (around 550 nm), for a wavelength range corresponding to the Bragg reflection peak, when propagation through the opal is (nearly) forbidden [6]. To be able to model the decrease in transmission for the shorter wavelengths, we have investigated various power-laws for the wavelength dependence of the absorption coefficient. We find a reasonable agreement for the transmission spectrum for an absorption coefficient resembling a Rayleigh-like scattering, $\alpha(\lambda) = \beta/\lambda^4$ with $\beta$ a constant. In all the following, we take $\beta$ = 1.5.$10^{-20}$ m$^3$, with the assumption that this coefficient is independent of the incidence angle. Remarkably, the Bragg transmission dip is satisfactorily reproduced by our calculation (fig. 8 a, b), relatively to its amplitude and width. This agreement persists when changing the side of irradiation for this transmission measurement (*i.e.* propagation through the substrate and the opal, or propagation through the opal and the substrate), or the polarization. When increasing the incidence angle (fig. 8 c, d), there is still a reasonable agreement between the experiment and the prediction, although we have not adjusted any parameter in the model, keeping the same parameter for the absorption (in the stratified model, this corresponds to a constant absorption per unit of "light traveled length", see appendix).

## 3. Fitting reflectivity measurements

Figure 9 presents the reflectivity spectra for a large range of incidence angles, TE and TM polarizations, and irradiation on the opal and substrate sides. The predictions of the model, in which the same set of parameters is used for all these experiments, are also provided. The choice of the explored $\lambda/D$ range implies the presence of a Bragg reflection peak, whose exact position should vary with the incidence angle. We observe indeed a very good agreement between the experimental and theoretical positions of these peaks, whose wavelength becomes smaller when increasing the incidence angle. In fig. 10, we have plotted the experimental peak wavelength for reflection as a function of the incidence angle. It is compared on the one hand with the predictions of our model, which includes both the periodical regions of the opal and the peripheral half layers, and on the other hand with the simplified Bragg model taking into account a global effective index (eq. 5). Although the differences remain marginal, our model seems more precise than the one with a global effective index.

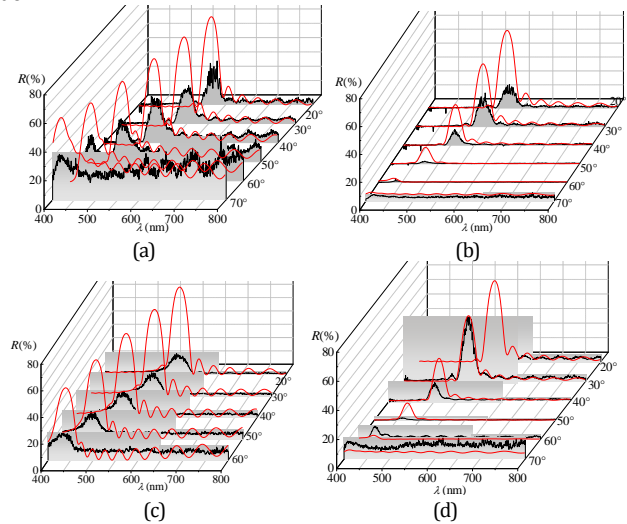

Fig. 9: Reflectivity spectra for various incidence angles $\theta$ (as indicated) for the same experimental sample (black) and related calculations (red on line) as in fig. 8: (a, b) on vacuum side; (c, d) on substrate side; with polarization (a, c):TE; (b, d): TM.

A specific added value of our model is that it predicts the width and amplitude of this Bragg peak, as it already did satisfactorily for transmission. Here, the estimates for the width appear in an acceptable agreement with the experiment. Our model is even able to produce sometimes an excellent fitting (*e.g.* 30° and 20° in TM) with the experiment, including a satisfactory agreement for the small oscillations, shown to be Fabry-Perot type (see section 3.C). Note that the exact thickness of our sample, and hence the phase of these oscillations, remains unknown because of the defects in the opal fabrication. These defects may also explain the observed quantitative discrepancies in the reflection amplitude. In particular, when the Bragg peaks appear smaller than predicted, it is natural to consider that the opal, with its successive layer-by-layer deposition, is not as periodical as in the ideal calculation. We also observe a satisfactory trend for the overall reflectivity (away from the Bragg peak) when varying the incidence angle, with differing behaviors for TE and TM polarizations. In particular, in TM polarization (fig. 9 b, d), the overall reflectivity decreases close to zero for large angles especially for irradiation from substrate side -*i.e.* large contrast between the input medium (substrate) and the first half-layer "gap"- . This is expected [29, 45] and is analogous to a near Brewster incidence angle, when the reflectivity undergoes a strong influence of the first half-layer.

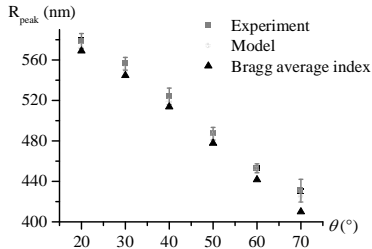

Fig. 10: Position of the peak of the reflectivity spectrum ($R_{peak}$) as a function of the incidence angle (substrate side). Comparison between the experiment, our calculations, and the simplified Bragg equation (sample and calculations are similar to fig. 9, TE polarization).

Finally, a more intriguing observation is on fig. 9d for $\theta = 50°$, where the residual experimental Bragg peak has a higher amplitude than predicted. This could originate in a residual depolarization of the reflected - or "backscattered"- light, or simply in an imperfect polarization of the light reaching opal. A marginal mixture of TE and TM polarizations inside the opal [36] would hence explain the observation of a Bragg peak when calculations rigorously limited to TM are not able to justify such an effect.

## 5. CONCLUSION AND PERSPECTIVES

To summarize, our simple model based on a stratified effective index in a direction $z$ perpendicular to the opal permits to evaluate the main features of reflection and transmission for on an opal deposited on a substrate, and allows for quantitative predictions. A major strength of such a one-dimensional model is that is well-suited to include the partial disorder of a real opal, particularly notable when fabricated by LB methods; oppositely, it cannot include the details of the crystalline structure of the opal with its various symmetry planes [*e.g.* the reflection on a (200) plane]. The need to introduce an *ad hoc* extinction coefficient to describe scattering may limit the range of validity of the quantitative predictions, because this coefficient is susceptible to vary strongly with the wavelength, or to a lesser extent with the incidence angle. Our experiments, despite their limited scope, show that these variations may be not so dramatic.

Our model, based upon a system of interferences between all the thin layers of the opal, makes it easy to discriminate between various physical effects, including interferences (Bragg peak at first and high-order, and Fabry-Perot type oscillations) with their wavy behaviors in reflection/transmission spectra and their sensitivity to a small change in the internal structure, as well as reflection from the interface region.

The demonstration of the influence of the first half layer, which intrinsically breaks the periodicity because of an interface with a substrate of an arbitrary index (or with vacuum), is an important result easily evidenced by our model. The size of the "gap" region, relatively to the wavelength, is an essential parameter. It must be emphasized that this corresponds to a very general situation, often unduly ignored when evaluating the optics of a photonic crystal. This "interface" coupling, occurring on a wavelength scale, easily hinders internal (in-depth) features, such as defects of the opal or of a photonic crystal.

The model that we have developed is highly flexible. It is easy to introduce a defective layer in the opal, and it would be also possible to introduce some dispersion in the average layer thickness, or to introduce an extra layer, non compactly arranged as can be imposed by the on-purpose [42] introduction of a special layer. Our one-dimensional model is intrinsically applicable to a compact arrangement of stacked parallel cylinders [21], provided that $f(z)$ is properly redefined in eq. 3. It can be also applied to an inverse opal. It is worth noting that although the effective medium theory is applicable when the inhomogeneities are much smaller than the wavelength, our quantitative approach uses very thin slices, or steps in the calculation, making our improved "effective index" model more applicable. Also, the orientation of the opal parallel to the substrate is an intrinsic source of structural anisotropy for polarization (see [32]; note also the differing TE/TM wavelength dependence in fig. 3), although all media are microscopically isotropic. These points have been discussed in literature [46]. In the same spirit, our formalism is applicable to an opal deposited on a prism, leading to the intriguing situation where propagation in the interface region (when $\lambda \leq D$) is mostly due to an evanescent wave.

At last, it is possible to consider the situation of an opal infiltrated by some material (liquid, gas, dopants), as if the infiltration is an added defect to the layer structure. This is done in an other work [32], where we calculate the optical response of a resonant material infiltrated in a photonic crystal, and demonstrate that under specific conditions, rather remote regions of the opal contribute to the reflectivity for well-chosen incidences.

## Acknowledgements

Work supported by the ANR project "Mesoscopic gas" 08-BLAN-0031. The opal was deposited at CRPP-Bordeaux in the Serge Ravaine group. We acknowledge discussions with Agnès Maitre group.

## **APPENDIX: The Matrix Transfer treatment**

We consider an incident light irradiation in the $z \leq 0$ region of index $n_0$ defined by its electric field:

$$\mathbf{E} = E_0 . \exp j(\omega t - \mathbf{k}.\mathbf{r}) \, \mathbf{u} \quad (A1)$$

with $\omega$ the (circular) frequency (corresponding to a wavelength in vacuum $\lambda$). In eq. (A1), $\mathbf{k}$ is the wave vector and $\mathbf{u}$ the unit vector, giving the direction of the polarization. As shown in fig. A1, the beam is incident under an incidence $\theta_0$ on a medium composed of N finite parallel layers (perpendicular to $z$, in the $z \geq 0$ region), and ended by a region of index $n_{N+1}$

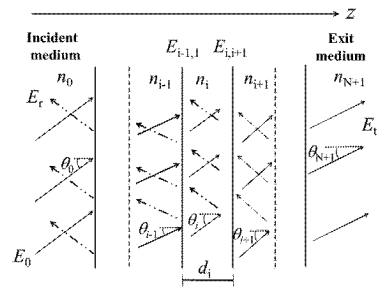

Fig. A1 : Schematics of the propagation in the stratified description. $E_{i,i+1}$ is the tangential component of the electric field at the interface between the $i^{th}$ layer and the $(i+1)^{th}$ layer. The direction of the plane waves (forward and backward) propagating in the $i^{th}$ layer is governed by the angle $\sin \theta_i = (n_0/n_i) \sin \theta_0$.

The matrix formalism takes into account the propagation in each layer, for the forward and backward field components resulting from successive transmission and reflection and which propagate under an angle satisfying the Snell's law at the successive interfaces.

Following the notations of fig. A1, one defines the tangential component of the electric (respectively magnetic) field at the generic boundary between the $(i\text{-}1)^{th}$ and $i^{th}$ layers as $E_{i\text{-}1,i}$ (respectively $H_{i\text{-}1,i}$). We first consider a transparent stratified medium. For the $i^{th}$ layer, the index is defined as $n_i$, the thickness as $d_i$, and the propagation angle as $\theta_i$ with:

$$n_0 \sin \theta_0 = n_i \sin(\theta_i) \qquad (A2)$$

The transfer matrix $M_i$ then appears when comparing the $(i\text{-}1)^{th}$ and $i^{th}$ layers boundary, with the $i^{th}$ and $(i+1)^{th}$ layers boundary. One finds:

$$\begin{pmatrix} E_{i-1,i} \\ H_{i-1,i} \end{pmatrix} = M_i \begin{pmatrix} E_{i,i+1} \\ H_{i,i+1} \end{pmatrix} \qquad (A3)$$

with:

$$M_i = \begin{pmatrix} \cos(\delta_i) & j\sin(\delta_i)/Y_i \\ jY_i\sin(\delta_i) & \cos(\delta_i) \end{pmatrix} \qquad (A4)$$

In eqs. (A3-A4), one has defined:

$$\delta_i = (2\pi/\lambda)\, n_i d_i \cos(\theta_i) \qquad (A5)$$

and $Y_i$ depends on the polarization:
for TE polarization:

$$Y_i = n_i \cos(\theta_i) \qquad (A6)$$

for TM polarization:

$$Y_i = n_i / \cos(\theta_i) \qquad (A7)$$

For N layers, the total matrix $M$ is the product of individual transfer matrices $M_i$, so that the input and output tangential fields can be calculated by multiplying the different transfer matrices together:

$$\begin{pmatrix} E_{0,1} \\ H_{0,1} \end{pmatrix} = M \begin{pmatrix} E_{N,N+1} \\ H_{N,N+1} \end{pmatrix} \qquad (A8)$$

with:

$$M = \begin{pmatrix} A & B \\ C & D \end{pmatrix} = M_1 M_2 \ldots M_N \qquad (A9)$$

The tangential components of the field at the input and output boundaries can be rewritten:

$$E_{0,1} = E_0 (1 + r) \qquad (A10)$$
$$E_{N,N+1} = t.E_0 \qquad (A11)$$
$$H_{0,1} = E_0 (1 - r) Y_0 \qquad (A12)$$
$$H_{N,N+1} = E_0\, t\, Y_{N+1} \qquad (A13)$$

with $r$ and $t$ the standard reflection and transmitted amplitude coefficients $r = E_r/E_0$ and $t = E_t/E_0$ and $E_r$ and $E_t$ defined in fig. A1

One hence deduces:

$$r = \frac{Y_0 A + Y_0 Y_{N+1} B - C - Y_{N+1} D}{Y_0 A + Y_0 Y_{N+1} B + C + Y_{N+1} D} \qquad (A14)$$

$$t = \frac{2 Y_0}{Y_0 A + Y_0 Y_{N+1} B + C + Y_{N+1} D} \qquad (A15)$$

from which the reflection and transmission intensity coefficients are simply calculated:

$$R = r.r^* \qquad (A16)$$

$$T = [n_{N+1} \cos(\theta_{N+1})/ n_0 \cos(\theta_0)].t.t^* \qquad (A17)$$

Phenomenological losses are introduced by an absorption coefficient for each slice ($\alpha_i$ for the $i^{th}$ slice) of the stratified index model, appearing in a complex index $N_i$:

$$N_i = n_i - j\, \kappa_i = n_i - j\alpha_i\, \lambda /4\pi \qquad (A18)$$

Note that it is only when the absorption integrated over a single layer of spheres remains small that the stratified model can be useful to describe the opal structure: indeed, a too strong scattering would make it very difficult to recognize an effect of the crystalline organization of the opal along $z$.

The previous matrix formalism still applies for media stratified with a complex index. The real angle $\theta_i$ is replaced by a complex value, noted $\theta'_i$ which no longer represents a direction of propagation [36, 38]. However, this complex incidence $\theta'_i$ still obeys the Snell's law [see eq. (A2)], and significantly, the quantity $N_i \sin(\theta'_i) = n_0 \sin \theta_0$ remains real because the input and output media (vacuum or glass) are transparent ($n_0$, $n_{N+1}$ are real). Moreover, the propagation direction remains unchanged if the absorption is weak ($\kappa_i \ll 1$). Indeed, the phase-term $\delta_i$ appearing in the matrices of eqs. (A3-A4), given without absorption by:

$$\delta_i = (2\pi/\lambda) n_i d_i \cos\theta_i = (2\pi/\lambda) d_i (n_i^2 - n_0^2 \sin^2\theta_0)^{1/2} \qquad (A19)$$

becomes complex with absorption:

$$\delta_i = (2\pi/\lambda) N_i d_i \cos\theta'_i = (2\pi/\lambda) d_i [(n_i - j\kappa_i)^2 - n_0^2 \sin^2\theta_0]^{1/2} \qquad (A20)$$

so that with the $\kappa_i \ll 1$ approximation, one has:

$$\delta_i \approx (2\pi/\lambda) d_i (n_i^2 - n_0^2 \sin^2\theta_0)^{1/2} - j(2\pi/\lambda)\, d_i\, \kappa_i n_i\, (n_i^2 - n_0^2 \sin^2\theta_0)^{-1/2} \qquad (A21)$$

which can be rewritten:

$$\delta_i \approx (2\pi/\lambda) d_i\, n_i \cos\theta_i - j(2\pi/\lambda)\, d_i\, \kappa_i n_i / \cos\theta_i$$
$$= \delta_i - j(2\pi/\lambda)\, d_i\, \kappa_i n_i / \cos\theta_i \qquad (A22)$$